\documentclass[aps,prl,twocolumn,epsfig]{revtex4}
\usepackage{graphics,graphicx,epsfig,color,hyperref,latexsym,float}

\begin{document}

\author{Rajarshi Tiwari and Pinaki Majumdar}

\affiliation{
  Harish-Chandra  Research Institute,\\
  Chhatnag Road, Jhusi, Allahabad 211019, India
}

\title{
The Crossover from a Bad Metal to a Frustrated Mott Insulator 
}

\date{20 Jan 2013}

\begin{abstract}
We use a novel Monte Carlo method to study the 
Mott transition in an anisotropic triangular 
lattice. The real space approach, retaining extended spatial
correlations, allows an accurate treatment of non trivial magnetic 
fluctuations in this frustrated structure.
Choosing the degree of anisotropy to mimic
the situation in the quasi-two dimensional organics,
$\kappa-$(BEDT-TTF)$_2$Cu[N(CN)$_2$]-X, we detect a wide 
pseudogap phase, with anomalous spectral and transport
properties, between the `ungapped' metal and the `hard gap' 
Mott insulator.
The magnetic fluctuations 
also lead to pronounced momentum dependence
of quasiparticle damping and pseudogap formation on the Fermi
surface as the Mott transition is approached.   
Our predictions about the `bad metal' state
have a direct bearing on the organics where they
can be tested via tunneling, angle resolved photoemission,
and magnetic structure factor measurement.
\end{abstract}

\maketitle

The Mott metal-insulator 
transition (MIT), and the proximity to a  Mott insulator
in doped systems, are crucial issues in correlated electron systems 
\cite{mott-rev0,mott-rev1,mott-rev2,mott-rev3}.
%
%
The Mott transition on a bipartite lattice is
now well understood, but the 
presence of triangular motifs in the structure
brings in geometric frustration
\cite{frust-rev1,frust-rev2}. 
This promotes  incommensurate
magnetic fluctuations whose nature, and
impact on the MIT, 
remain outstanding problems.

The organic salts provide a concrete testing
ground for these effects
\cite{org-rev-kanoda,org-rev-mack}.
The 
$\kappa-$(BEDT-TTF)$_2$Cu[N(CN)$_2$]-X 
salts are quasi two dimensional (2D) 
materials where the BEDT-TTF 
dimers define a triangular
lattice with anisotropic hopping 
\cite{org-struct-hopping}. 
The large lattice spacing, $\sim 11 \AA$,
leads to a low bandwidth, enhancing electron correlation 
effects, while the triangular motif disfavours Neel
order. 
The X$=$Cl$_{1-x}$Br$_x$ 
family shows a MIT as $x$ drops
below $\sim 0.75$ \cite{org-ph-diag}.
The metallic state is {\it very incoherent}
above $T \sim 50$K: 
the resistivity \cite{org-res} is large, 
$\gtrsim 100 $m$\Omega$cm,
the optical response has non Drude 
character \cite{org-opt1,org-opt2}, 
and NMR \cite{org-nmr1,org-nmr2} suggests
the presence of a pseudogap (PG).
How these properties arise in response to
magnetic fluctuations, and the crucial low energy
spectral features in the vicinity of the Mott transition,
remain to be clarified.

We use a completely new approach to the Mott transition,
using auxiliary fields, that emphasizes the
role of spatial correlations near the MIT.
Our principal results, based on Monte Carlo (MC) on
large lattices are the following.
(i)~The interaction $(U)$-temperature $(T)$ phase
diagram that we establish has a striking correspondence
with $\kappa$-BEDT in terms of
magnetic transition and re-entrant insulator-metal 
crossovers.
(ii)~At intermediate temperature, in the
magnetically disordered regime,
we obtain a strongly non Drude optical response
in the metal,
and predict a pseudogap (PG) phase over a wide
interaction and temperature window.
(iii)~The electronic spectral function $A({\bf k}, \omega)$
is {\it anisotropic}
on the Fermi surface, with both the damping rate and
PG formation showing a clear angular dependence
arising from coupling to
incommensurate magnetic fluctuations.

\begin{figure}[b]
\centerline{
\includegraphics[width=6.8cm,height=5.0cm]{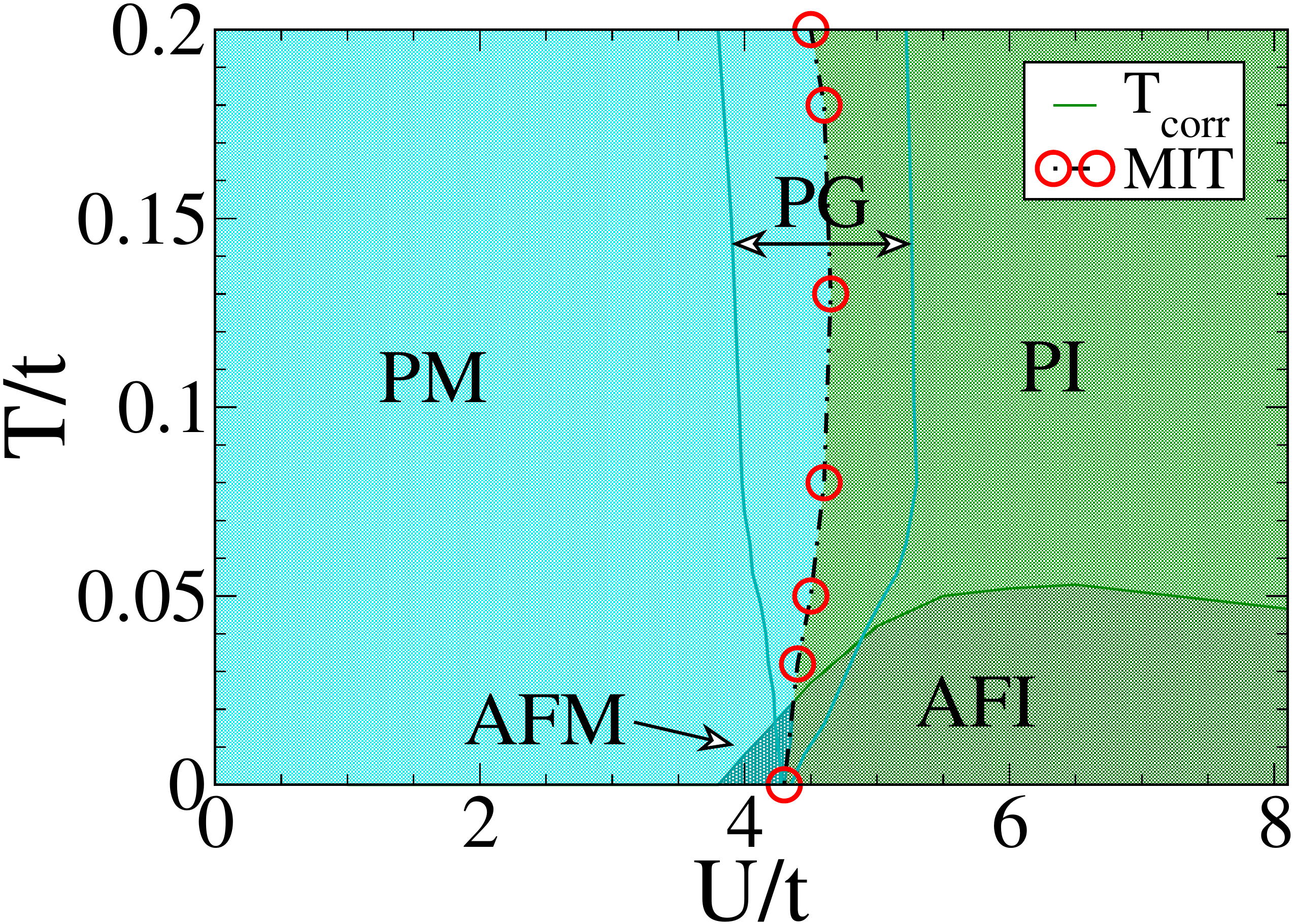}
}
\caption{\label{phd}
$U-T$ phase diagram of the Hubbard model at $t'/t=0.8$.
The phases are paramagnetic metal (PM), paramagnetic insulator (PI),
antiferromagnetic metal (AFM) and antiferromagnetic insulator (AFI).
The AFM and AFI are not simple Neel ordered. PG indicates a
pseudogap phase, metallic or insulating.
There is no genuine magnetic transition in two dimensions so our
$T_{corr}$ indicates the temperature where the magnetic correlation
length becomes larger than the system size $24 \times 24$.
The MIT line 
is determined from change in sign of the temperature
derivative of resistivity, {\it i.e}, $d\rho/dT =0$.
}
\end{figure}

To provide a quick background, there have been several studies of 
the single band Hubbard model on a triangular lattice\cite{th-hubb-Imada,th-hubb-Ino,th-hubb-Kawa,th-hubb-Becca,th-hubb-Lime,th-hubb-Scale,th-hubb-Phillips,th-hubb-Kotl1,th-hubb-Ohashi,th-hubb-Liebsch,th-hubb-Sato,th-hubb-Senech,th-hubb-Mila}
to model organic physics.
Dynamical mean field theory (DMFT)
has been the method of choice
\cite{th-hubb-Lime,th-hubb-Scale,th-hubb-Phillips}, usually used in its
cluster variant (C-DMFT) 
\cite{th-hubb-Kotl1,th-hubb-Ohashi,th-hubb-Liebsch,th-hubb-Sato}
to handle short range spatial correlations.
The results depend on the degree of frustration
and the specific method
but broadly suggest  the following:
(i) the ground state is a PM Fermi liquid 
at weak coupling, a `spin liquid' PI 
at intermediate coupling, and 
an AFI 
at large coupling\cite{th-hubb-Imada,th-hubb-Ino,th-hubb-Kawa,th-hubb-Becca},
(ii) the qualitative features in optics  \cite{org-opt1}
and transport \cite{th-hubb-Lime} are recovered, 
(iii) there could be 
a re-entrant insulator-metal-insulator transition 
with increasing 
temperature for a certain window of frustration
\cite{th-hubb-Ohashi,th-hubb-Liebsch},
(iv) the low temperature 
SC state could emerge 
\cite{th-hubb-org-sc1,th-hubb-org-sc2,th-hubb-org-sc3}
from  Hubbard physics, although there is no
consensus~\cite{th-hubb-org-sc-maz}.

Surprisingly, there seems to be limited effort
on the nature of {\it spatial fluctuations}, which
could be significant in this low
dimensional frustrated system.
To clarify this aspect we study the single band Hubbard model on the 
anisotropic triangular lattice:
\begin{equation}
H=\sum_{\langle ij\rangle\sigma}t_{ij} 
c^{\dagger}_{i\sigma}c_{j\sigma} -\mu\sum_i n_i
  + U\sum_{i}n_{i\uparrow}n_{i\downarrow}
\end{equation}
We use a square lattice geometry  but with the following
anisotropic hopping: 
$t_{ij}=-t$ when ${\bf R}_i - {\bf R}_j = \pm {\hat x}a_0$
or $\pm {\hat y}a_0$, where $a_0$ is the lattice spacing, and
$t_{ij}=-t'$ when ${\bf R}_i - {\bf R}_j = \pm({\hat x} + {\hat y})
a_0$. 
We will set $t=1$ as the reference energy scale.
$t'=0$ corresponds to the square lattice, and $t'=t$ to
the isotropic 
triangular lattice. We have studied the problem over the
entire $t'/t$ window $[0,1]$, but focus on $t'/t=0.8$ in
this paper. 
$\mu$ controls the electron density, which we maintain 
at half-filling, $n=1$. $U >0$ is the Hubbard repulsion.

We use a Hubbard-Stratonovich (HS) transformation
that introduces a vector field ${\bf m}_i(\tau)$ and a scalar
field $\phi_i(\tau)$ at each site 
\cite{vec-hubb-hs1,vec-hubb-hs2}
to decouple the interaction.
We need  two approximations 
to make progress. (i)~We will treat the ${\bf m}_i$ and $\phi_i$
as classical fields, {\it i.e}, neglect their time dependence.
(ii)~While we completely retain the thermal fluctuations in
${\bf m}_i$, we treat $\phi_i$ at the saddle point level, 
{\it i.e}, $\phi_i \rightarrow \langle \phi_i \rangle = (U/2)
\langle \langle n_i \rangle \rangle = U/2$ at half-filling.
With this approximation the half-filled problem is 
mapped on to electrons coupled to the field ${\bf m}_i$
(see Supplement). 
\begin{equation}
H_{eff} 
=\sum_{ij,\sigma}t_{ij} c^{\dagger}_{i\sigma}c_{j\sigma}
- {\tilde \mu} N 
- \frac{U}{2}\sum_{i}{\bf m}_{i}\cdot\vec{\sigma}_{i}
+ \frac{U}{4}\sum_{i}{\bf m}_{i}^{2}
\end{equation}
where ${\tilde \mu} = \mu -U/2$.
We can write $H_{eff} = H_{el}\{{\bf m}_i\} + H_{cl}$,
where $H_{cl}= (U/4)\sum_i {\bf m}_i^2$.
For a {\it given} configuration $\{{\bf m}_i\} $ one just needs 
to diagonalise $H_{el}$, but the $\{{\bf m}_i\} $ themselves
have to be determined from the distribution
\begin{equation}
P\{{\bf m}_i\} = {{ 
Tr_{c,c^{\dagger}} e^{-\beta H_{el} } e^{-\beta H_{cl}}
} 
\over
\int {\cal D}{\bf m} 
Tr_{c,c^{\dagger}} e^{-\beta H_{el} } e^{-\beta H_{cl}}
}
\end{equation}
Equation (2)  
describes electron propagation in the
${\bf m}_i$ background, while equation (3) describes how the
${\bf m}_i$ emerge and are spatially correlated due to
electron motion.
The neglect of dynamics in the ${\bf m}_i$ reduces the 
method to unrestricted Hartee-Fock (UHF) 
mean field theory at $T=0$. However, the exact 
inclusion of classical thermal fluctuations quickly
improves the accuracy of the method with
increasing temperature. We will discuss the limitations
of the method further on.

\begin{figure}[b]
\centerline{
\includegraphics[width=7.0cm,height=6.3cm]{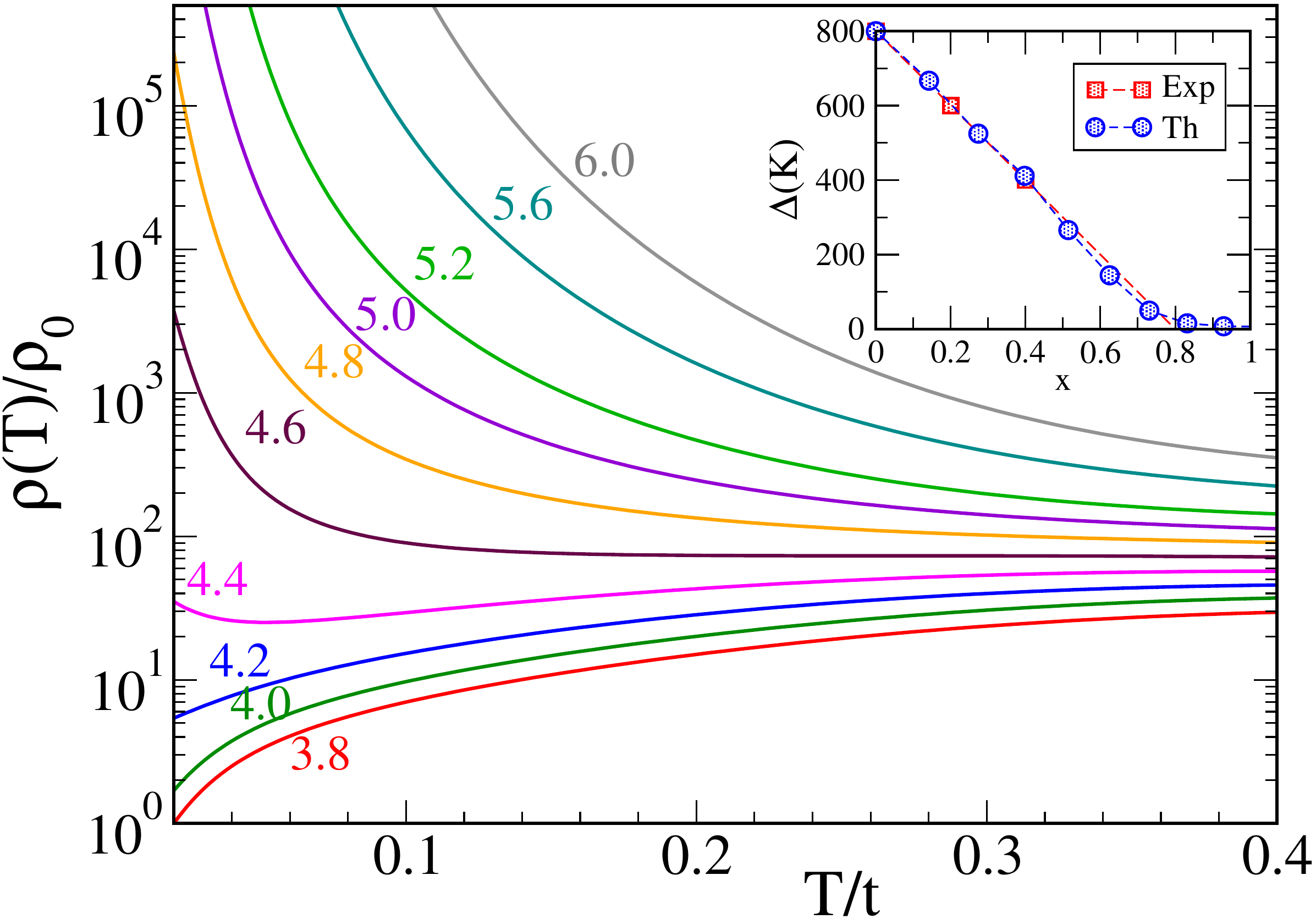}
}
\caption{\label{rho}
Temperature dependence of the resistivity for different
$U$ in the neighbourhood of $U_c$. The normalising scale 
is $\rho_0 = {{\hbar} c_0}/{\pi e^2}$ (see text). This is 
$\sim 380 \mu \Omega$cm for the organics. 
The $U/t$ values are indicated on the curves.
The inset shows the experimental transport gap in the 
Cl$_{1-x}$Br$_x$ family (in Kelvins), 
and our estimated transport gap. We used a fit
$U(x)/t = 6 - 1.35x -0.4x^2$ (see text) to 
reproduce the transport gap\cite{org-res} estimated from 
the experiments.
}
\end{figure}

Due to the fermion trace,
$P\{{\bf m}_i\}$ is not exactly calculable.
To generate the equilibrium $\{{\bf m}_i\} $
we use MC sampling
\cite{hs-mc-1,hs-mc-2,hs-mc-3}. 
Computing the energy cost of
an attempted update requires  diagonalising $H_{el}$.
To access large sizes within limited time, we 
use a cluster algorithm \cite{tca}
for estimating the update cost.
Rather than diagonalise the full $H_{eff}$ for every 
attempted update, we calculate the energy cost of an update by
diagonalizing a small cluster (of size $N_c$, say) around 
the reference site. We have extensively benchmarked 
this cluster based MC method\cite{tca}.
The MC was done
for lattice of size $N=24 \times 24$, with clusters of size 
$N_c=8 \times 8 $. We
calculate the thermally averaged  structure factor
$S({\bf q}) = \frac{1}{N^2}\sum_{ij}\langle{\bf m_{i}}
\cdot{\bf m_{j}}\rangle e^{i{\bf q}\cdot({\bf R}_i-{\bf R}_j)}$
at each temperature. The onset of rapid growth in $S({\bf q})$
at some ${\bf q} = {\bf Q}$, say, indicates a magnetic
transition. The electronic properties are calculated
by diagonalising $H_{el}$ on the full lattice for equilibrium
$\{{\bf m}_i\} $ configurations.

Fig.\ref{phd} shows the $U-T$ phase diagram  at $t'/t=0.8$.
Our low temperature result is equivalent to UHF and leads 
to a transition from an uncorrelated paramagnetic metal to
an incommensurate AF metal with wavevector 
${\bf Q} = {\bf Q}_1 \sim \{0.85 \pi,
0.85 \pi\}$ at $U_{c1} \sim 4.0t$. At $U_{c2}
\sim 4.5t$ there is a transition to an AF `Mott' 
insulator
with ${\bf Q}_2 \sim \{0.8\pi,0.8\pi\}$. 
The magnitude  $m_i=\vert {\bf m}_i \vert$ 
is small in the AFM and grows
as $U/t$ increases in the Mott phase. 
The existence  of the AF metal, and the nature of
order in the intermediate $U/t$ Mott phase,  could be 
affected by the neglected 
quantum fluctuations of the ${\bf m}_i$. 

Finite temperature brings into play the low energy fluctuations 
of the ${\bf m}_i$. The effective model has the $O(3)$ symmetry
of the parent Hubbard model so it cannot sustain true long
range order at finite $T$. However, our annealing results suggest
that magnetic correlations grow rapidly below a temperature
$T_{corr}$, and weak interplanar coupling would stabilise in plane
order below $T_{corr}$.
This scale increases from zero at $U = U_{c1}$, reaches 
a peak at $U/t \sim 6.5$, and falls beyond as the
virtual kinetic energy gain reduces with
increasing $U$. 

We classify the finite $T$ phases as metal or insulator based 
on $d\rho/dT$, the temperature derivative
of the resistivity. The dotted line indicating the MIT corresponds
to the locus $d\rho(T,U)/dT=0$.
In addition to the magnetic and transport classification 
we also show a window around the MIT line where the 
electronic density of states 
(DOS) has a pseudogap. To the
right of this region the 
DOS has a `hard gap' while to 
the left it is featureless. The MIT line shows re-entrant 
insulator-metal-insulator behavior with
increasing $T$ near $U \sim U_{c2}$.

We can attempt a quick comparison of the phase diagram 
with that in the $\kappa$-BEDT family. The primary hopping 
is $t \sim 65$meV, and $t'/t \sim 0.8$
\cite{org-struct-hopping}
(a recent
{\it ab initio} estimate suggests $t'/t \lesssim 0.5$
for $\kappa$-Cl).
Fitting the transport gap in $\kappa$-Cl
(see Fig.\ref{phd} inset)
suggests that $U/t \sim 6.0$ at $x=0$. 
From our results
this would indicate that $T_{corr}/t
\sim 0.05$, at $x=0$,
{\it i.e}, $T_{corr} \sim 35$K, 
not too far from the NMR inferred $T_c
\sim 30$K. 

Fig.\ref{rho} shows the resistivity $\rho(T)$ computed via
the Kubo formula for varying $U/t$. 
We first compute the planar resistivity (which has the
dimension of resistance) and then compute the effective
three dimensional resistivity of `decoupled' layers (see
Supplement) by using the $c$-axis spacing, $c_0$. 
In the  Cl$_{1-x}$Br$_x$ family it is
observed that the transport gap can be
fitted to $\Delta(x) 
\approx 800 -1000x$ Kelvin\cite{org-res}. 
We match this to the $U$
dependence of our calculated gap, $\Delta(U)/t$,
and infer $U/t\vert_{x=0} \sim 6$. The MIT occurs at $x_c
\sim 0.75$ and for us at $U/t \approx 4.5$. 
Fitting a quadratic form to $U(x)/t$ to capture the 
measured transport gap,
we get $U(x)/t \approx 6-1.35x-0.4x^2$. The $U/t$ range in Fig.\ref{rho} 
corresponds roughly to $x=[0,1]$. Since $t=65$meV,
$T=0.4t$ is approximately $300$K. 
\begin{figure}[b]
\centerline{
\includegraphics[width=8.2cm,height=4.0cm]{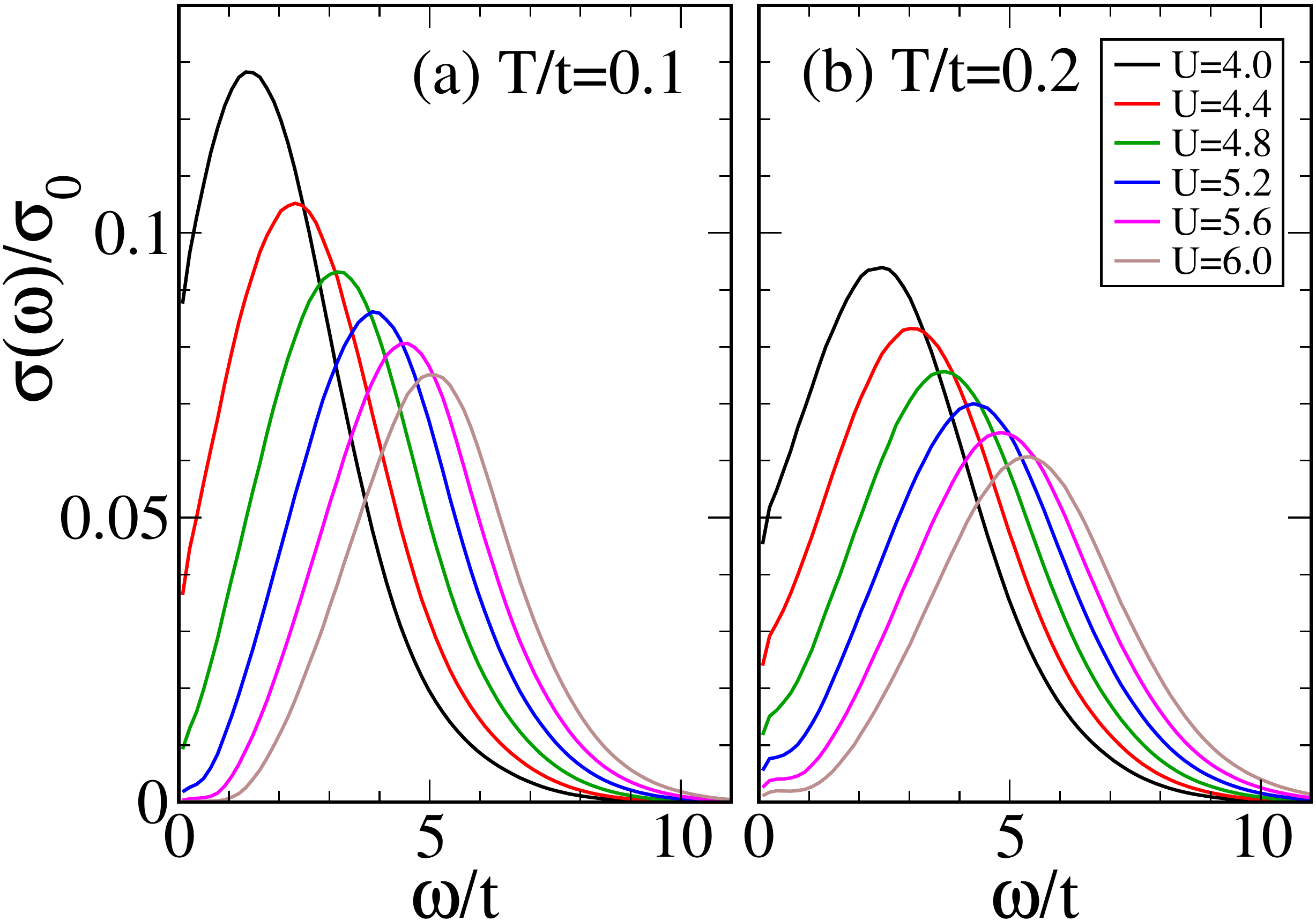}
}
\caption{\label{optics}
Optical conductivity at $T/t=0.1$ and $0.2$ for $U$
varying across $U_c$.
At these temperatures the $\sigma(\omega)$ is non Drude even
in the `weakly correlated' case $U/t \sim 4.0$. The finite
frequency peak evolves into the Hubbard transition at large
$U/t$. For a rough comparison to organics, $T/t=0.1
\equiv $80K, $\omega/t =5 
\equiv  2500$cm$^{-1}$, and $\sigma/\sigma_0 =0.1 \equiv
265\Omega^{-1}$cm$^{-1}$.
}
\end{figure}

Our resistivity is in units of
$\rho_0=\frac{\hbar c_0}{\pi e^2}$.
Using $c_0\sim 29\AA$, $\rho_0\sim 380\mu\Omega$cm, yields $\rho\sim 60\rho_0\sim 25$m$\Omega$cm
at $T \sim 0.4t$, while experimental value is $\gtrsim 100$m$\Omega$cm.
The difference could come from electron-phonon scattering absent in our model. 
Limelette {\it et al}\cite{th-hubb-Lime} presented DMFT based resistivity result
that compares favourably with experiments, but, apparently, involves
an arbitrary scale factor. Our re-entrant window $\delta U$=0.4$t$ near $U_c$, 
inferred from thermally driven I-M-I crossover, is equivalent to $\delta x$=0.2.
This is consistent with $\delta x\sim 0.2$ in the Cl$_{1-x}$Br$_x$ 
family\cite{org-res}. The C-DMFT estimates of the re-entrant window varies 
widely, from $\delta U$$\sim$0.3$t$\cite{th-hubb-Liebsch} 
to $\sim$1.2$t$\cite{th-hubb-Ohashi}.
\begin{figure}[t]
\centerline{
\includegraphics[width=8.4cm,height=4.8cm]{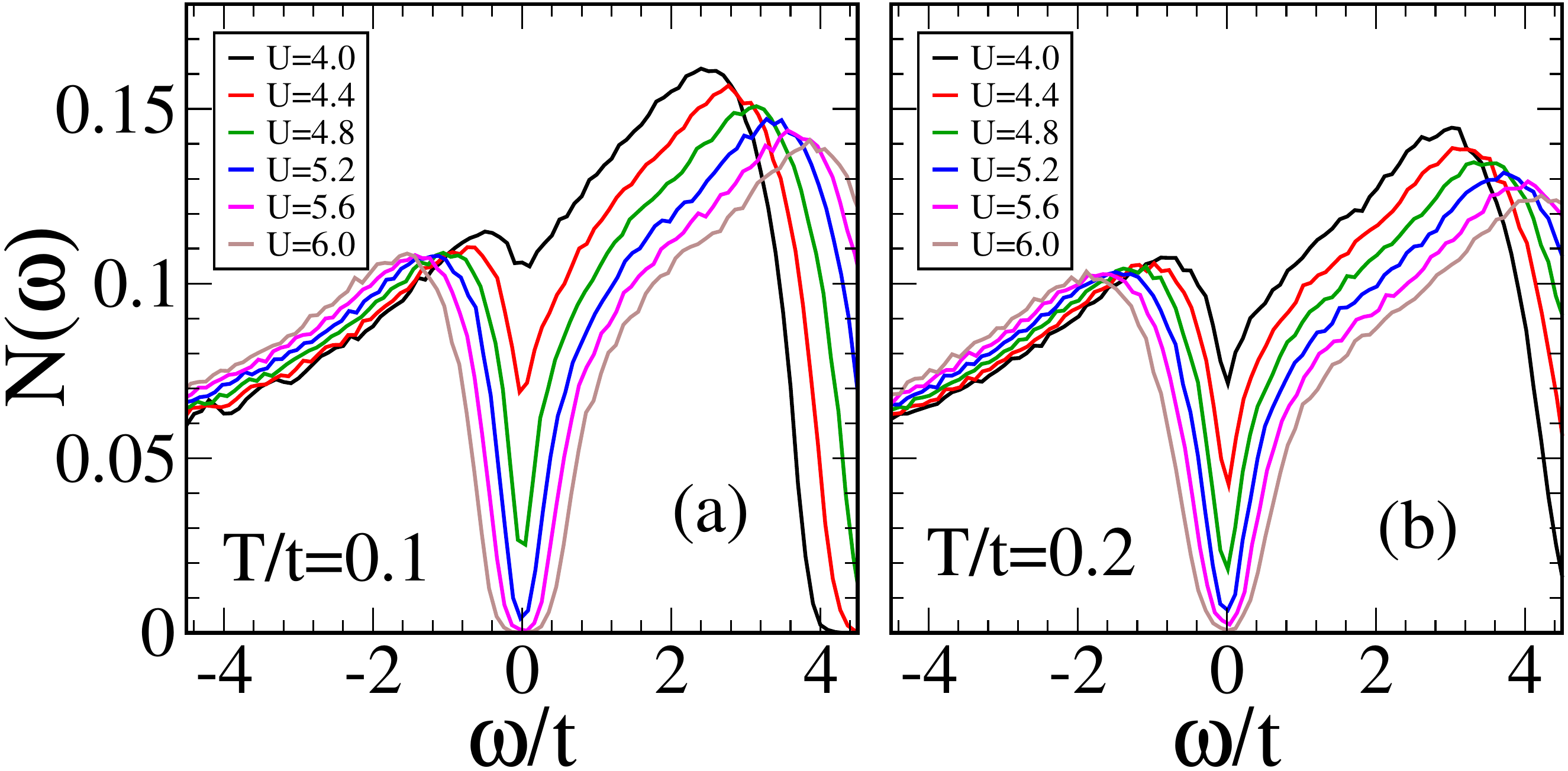}
}
\caption{\label{dos}
Density of states at $T/t=0.1,~0.2$ for $U$ 
varying across $U_c$. The dip in the DOS deepens with 
increasing $T$ for $U/t \lesssim 4.8$. For larger $U/t$ 
the system slowly gains spectral weight with 
increasing $T$.
}
\end{figure}

\begin{figure*}[t]
\centerline{
\includegraphics[width=15.8cm,height=3.0cm]{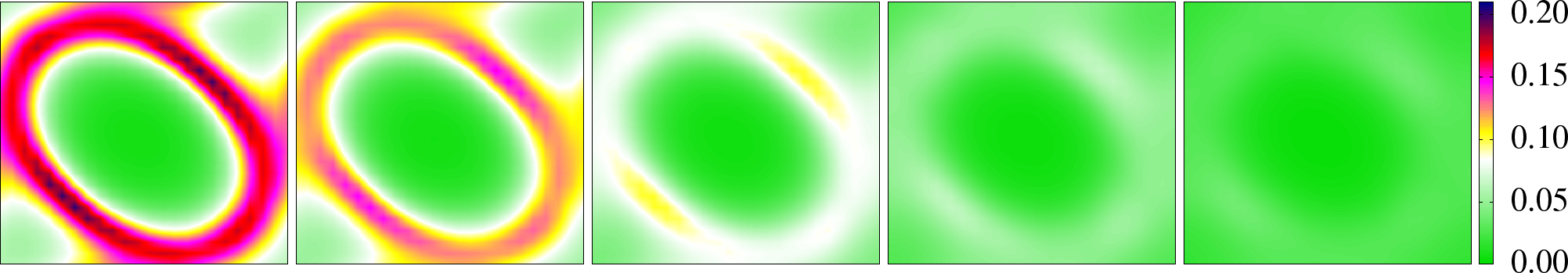}\hspace{0.2cm}
}
\vspace{.2cm}
\centerline{
\includegraphics[width=16.0cm,height=3.0cm]{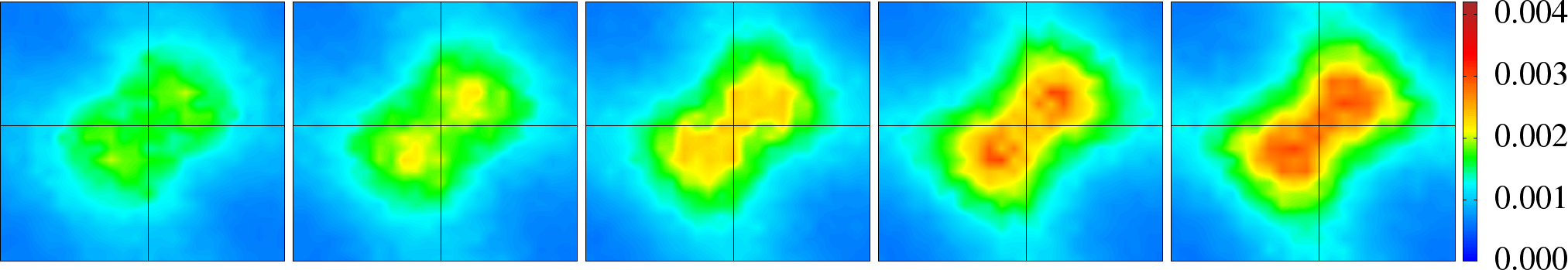}
}
\caption{\label{aksq}Top:~
Momentum dependence of the low frequency spectral weight in
the electronic spectral function $A({\bf k}, \omega)$ at
$T/t=0.1$.
$k_x,k_y$ range from [-$\pi,\pi$] in the panels. Note the
systematically larger weight near ${\bf k} = [\pi/2,\pi/2]$ and
$[-\pi/2,-\pi/2]$ and smaller weight in the segments near
$[\pi,0]$ and $[0,\pi]$.
$U/t=4.2,~4.4,~4.6,~4.8,~5.0$, left to right.
Bottom:~
Magnetic structure factor $S({\bf q})$ for the
auxiliary fields ${\bf m}_i$ for the same set of $U/t$.
The $q_x,q_y$ range from $[0,2\pi]$. Note the very weak and
diffuse structure at $U/t=4.2$ and the much larger and 
differentiated structure at $U/t=5.0$.
}
\end{figure*}

Fig.\ref{optics} shows the optical conductivity $\sigma(\omega)$ at
$T =0.1t$ and $T=0.2t$ as $U/t$ varied across the Mott
crossover.  Our first observation is the
distinctly non Drude nature of $\sigma(\omega)$ in
the metal, $U/t \lesssim 4.4$, with 
$d\sigma(\omega)/d\omega\vert_{\omega \rightarrow 0} 
> 0$. The low frequency hump in the bad metal evolves
into the interband Hubbard peak in the Mott phase.
The  change in the lineshape with increasing $T$ is
more prominent in the metal, with the peak location
moving outward, and is more modest deep in the
insulator.

In the organics 
the experimenters have carefully isolated 
the Mott-Hubbard features in the spectrum by
removing phononic and intra-dimer effects \cite{org-opt2}.
Since we have already fixed our $t,t',U$
we have no further fitting parameter for $\sigma(\omega)$.
The measured spectrum at $x \sim x_c$ and $T \sim 50-90$K
has a peak around $1500-2000$cm$^{-1}$.
Using $U_c/t \sim 4.5$ and $T/t =0.1$ we get 
$\omega_{peak}/t \sim 3.0$, which translates to
$\sim 1500 $cm$^{-1}$. 
The  magnitude of our $\sigma(\omega)$ at
$\omega_{peak}$ is $\sim  0.1 \sigma_0
\sim 265 \Omega^{-1}$cm$^{-1}$, since $\sigma_0 = 1/\rho_0
\sim 2650 \Omega^{-1}$cm$^{-1}$. This is remarkably close 
to the measured value, 
$\approx 280 \Omega^{-1}$cm$^{-1}$(Ref.~\cite{org-opt2} Fig.3).

While the characteristic scales
in $\sigma(\omega)$ match well 
with experiments, the experimental spectrum has 
weaker dependence on temperature
and composition $x$. This could
arise from the subtraction process and the presence
of other interactions in the real material. Our result
differs from DMFT \cite{org-opt1}, 
and agrees with the experiments, in that
we do not have any feature at $\omega =U/2$. We have verified the
{\it f}-sum rule numerically.

The crossover from the bad metal
to the insulator involves a wide window with a pseudogap
in the electronic DOS, $N(\omega)$. One may have guessed this from
the depleting low frequency weight in $\sigma(\omega)$, Fig.\ref{dos}
makes this feature explicit. We are not aware of
tunneling studies in the organics, but our results
indicate a wide window, $U/t \sim [4,5.3]$, where there
is a distinct pseudogap in the global DOS.
This suggests that the entire $x \sim [1.0,0.35]$
window in the organics should have a PG.
For $U/t \lesssim 4.6$ the dip feature deepens with
increasing $T$, we have $dN(0)/dT <0$ (compare panels (a) and (b), Fig.\ref{dos}),
while for $U/t \gtrsim 4.6$ we have a weak $dN(0)/dT >0$. 
The PG arises from the coupling of electrons to the fluctuating ${\bf m}_i$.
A large $m_i$ at all sites would open a Mott gap, independent
of any order among the moments. Weaker $m_i$, thermally
generated in the metal  near $U_{c1}$ 
and with only short range correlations, manages to deplete
low frequency weight without opening a gap. Since the
typical size $\langle m_i \rangle$ increases with $T$
in the metal, we see the dip deepening at $U<U_c$.

While the {\it size} of the ${\bf m}_i$ determine the
overall depletion of DOS near $\omega=0$ and the opening
of the Mott gap, the {\it angular correlations} dictate
the momentum dependence of the spin averaged 
electronic spectral function $ A({\bf k},\omega)$ (see
Supplement).

Within `local self energy' picture,
as in DMFT, $A({\bf k},\omega)$ should
have no ${\bf k}$ dependence on the Fermi surface (FS).
In that case we should have ${\bf k}$ independent
suppression of $A({\bf k}, 0)$ with increasing $U/t$.

Fig.\ref{aksq}, top row,  shows maps of $A({\bf k},0)$
for $k_x,k_y =[-\pi,\pi]$, at $T/t=0.1$,
as increasing $U/t$ transforms the 
bad metal to a Mott insulator. The first panel at
$U/t=4.2$ (roughly a Br sample) shows
weak anisotropy  on the nominal FS while 
Fig.\ref{dos}.(a) suggests that a weak PG has already 
formed.  At $U/t=4.4$, next panel, 
the weak anisotropy is 
much amplified and the weight in the
$[0,0]\rightarrow [\pi,\pi]$ direction
is distinctly larger. 
%
%
 The next three panels basically 
show insulating states but without a hard Mott gap.
Overall, the `destruction' of the FS seems to
start near $[\pi,-\pi]$, the `hot' region, and 
ends with the region near $[\pi/2,\pi/2]$, the `cold spot'.
We show data on the full $A({\bf k},\omega)$ in the
Supplement that indicates that with increasing $U$ a 
PG feature forms at the hot spot while the cold spot
still has a quasiparticle peak.

Second row in Fig.\ref{aksq} shows the 
$S({\bf q})$ of the auxiliary fields at $T/t=0.1$
for the same $U/t$ as in the upper row. While there is
no magnetic order we can see the growth of correlations
centered around ${\bf Q} \approx [0.85 \pi, 0.85\pi]$ 
as $U/t$ increases. The dominant electron scattering
would be from ${\bf k}$ to ${\bf k} + {\bf Q}$, and 
the impact would be greatest in regions of the
FS in the proximity of minima in 
$\vert \nabla \epsilon_{\bf k} \vert $.
The location of the hot spots on the FS,
and the momentum connecting them, indeed 
correspond to this scenario. 

While we have a method that captures 
non trivial spatial correlations and its impact on
electronic properties, we need to be cautious about
some shortcomings. (i)~Neglecting the dynamics of
the ${\bf m}_i$ misses correlation effects in the
ground state of the metal and underestimates $U_c/t$.
(ii)~It also misses the `Fermi liquid' physics in the 
low $T$ metal, but should be reliable in the 
$T/t \gtrsim 0.1$ regime that we have focused on.
(iii)~There is potentially a `spin liquid' insulator
\cite{th-hubb-Senech,th-hubb-Mila} at intermediate 
$U/t$ and $t'/t=1$.
We do not know of such results at $t'/t=0.8$, but 
would prefer to emphasize our finite $T$ results rather
than the nature of the ground state.

{\it Conclusion:}
We introduced and explored in detail a method which
retains the spatial correlations that are crucial near the 
Mott transition on a frustrated lattice.
Using electronic parameters that describe 
the $\kappa$-BEDT 
based organics we obtain a magnetic $T_c$, 
metal-insulator phase diagram, and optical
response that reproduces the key experimental scales.
We uncover a wide pseudogap regime near the MIT, and 
predict distinct signatures of the incommensurate magnetic 
fluctuations in the angle resolved photoemission
spectrum.

We acknowledge use of the HPC clusters at HRI. PM acknowledges support
from the DST India (Athena), a DAE-SRC Outstanding Research
Investigator grant, and a discussion with T. V. Ramakrishnan.

\bibliographystyle{unsrt}

\newpage
\section{Supplementary information:}

\subsection{Derivation of the effective Hamiltonian}
Our starting point is the Hubbard model
\begin{eqnarray}
H &=& \sum_{\langle ij\rangle\sigma}t_{ij} 
c^{\dagger}_{i\sigma}c_{j\sigma} -\mu\sum_i n_i
  + U\sum_{i}n_{i\uparrow}n_{i\downarrow} \cr
&=& H_0 -\mu\sum_i n_i 
+ U\sum_{i}n_{i\uparrow}n_{i\downarrow} \nonumber
\end{eqnarray}
We implement a rotation invariant decoupling of
the Hubbard interaction as follows.
First, one can write
$$
n_{i\uparrow}n_{i\downarrow}=
\frac{n_i^2}{4}-(\vec{s}_i\cdot {\hat {\bf m}_i})^2
$$
where $n_i =n_{i\uparrow}+n_{i\downarrow}$ is the charge 
density, 
$\vec{s}_{i}=\frac{1}{2}\sum_{\alpha,\beta}
c^{\dagger}_{i\alpha}\vec{\sigma}_{\alpha\beta}c_{i\beta}=2\vec{\sigma}_i$
is the local electron spin operator, and ${\hat {\bf  m}_i}$ is an
{\it arbitrary unit vector}.

The partition function of the Hubbard model is
\begin{eqnarray}
Z & = & \int D[c,\bar{c}]e^{-S} \cr
S & =&  \int_{0}^{\beta}d \tau {\cal L}(\tau) \cr
{\cal L} & = & 
\sum_{i\sigma}
\bar{c}_{i\sigma}(\tau)\partial_{\tau}c_{i\sigma}(\tau)
+ H(\tau) 
\nonumber
\end{eqnarray}
We can introduce two space-time varying auxiliary fields
for a Hubbard-Stratonovich transformation:
(i)~$\phi_{i}(\tau)$ coupling to charge density, and 
(ii)~$\Delta_{i}(\tau) {\hat {\bf m}_i}(\tau)={\bf m}_i(\tau)$
coupling to electron spin density ($\Delta_i$ is real positive).
This allows us to define a 
SU(2) invariant HS transformation (see ref.~\cite{weng,dupuis}),
$$
  e^{U n_{i\uparrow}n_{i\downarrow}}  = 
  \int \frac{d\phi_{i} d {\bf m}_i}{4\pi^{2}U}
  e^{ 
\left( 
\frac{\phi_{i}^2}{U}+i\phi_{i} n_i
+ \frac{\bf {m}_i^2}{U}-2{\bf m}_i\cdot\vec{s}_{i}
\right)
}
$$
The partition function now becomes:
\begin{eqnarray}
  Z &=& 
\int\prod_{i}\frac{d\bar{c_{i}}dc_{i}d\phi_{i} d{\bf m}_{i}}{4\pi^{2}U}
  e^{\left(-\int_{0}^{\beta}{\cal L}(\tau)\right)} \cr
{\cal L}(\tau) &=& 
\sum_{i\sigma}
\bar{c}_{i\sigma}(\tau)\partial_{\tau}c_{i\sigma}(\tau)
+ H_0(\tau)
+ {\cal L}_{int}(\phi_i(\tau), {\bf m}_i(\tau)) \cr
{\cal L}_{int} &=& \sum_{i}\left[\frac{\phi_{i}^2}{U}+
i\phi_{i} n_i+
 {{\bf m}_i^2 \over U}  - 2 {\bf m}_{i}\cdot\vec{s}_{i}\right]
\nonumber
\end{eqnarray}

As discussed in the text, to make progress we need two 
approximations: (i)~neglect the time ($\tau$) dependence of the 
HS fields, (ii)~replace the field $\phi_i$ by its
saddle point value $(U/2)\langle n_i \rangle =U/2$,
since the important low energy fluctuations arise
from the ${\bf m}_i$. 
Substituting these, and simplifying the action, one gets the 
effective Hamiltonian 
$$
H_{eff} = H_{0} - {\tilde \mu} \sum_i n_i 
- \sum_{i} {\bf m}_i\cdot\vec{\sigma}_{i}
+ \sum_{i} {{\bf m}_i^2 \over U} 
$$
where ${\tilde \mu} = \mu - U/2$.
For convenience
we redefine ${\bf m}_i  \rightarrow \frac{U}{2} {\bf m}_i $, so that 
the ${\bf m}_i$ is dimensionless.
This leads to the effective Hamiltonian used in the text:
$$
H_{eff} = H_{0} - {\tilde \mu} \sum_i n_i 
- {U \over 2} \sum_{i} {\bf m}_i\cdot\vec{\sigma}_{i}
+ {U \over 4} \sum_{i} {\bf m}_i^2  
$$
The partition function can be written as:
$$
Z = \int {\cal D} {\bf m}_i Tr_{c,c^{\dagger}} e^{-\beta H_{eff}}
$$
For a given configuration $\{ {\bf m}_i\}$ 
the problem is quadratic in the fermions,
while the configurations themselves are obtained by a Monte
Carlo as discussed in the text.

\subsection{Optical conductivity}
The conductivity of the two dimensional system is first
calculated as follows 
(ref.\cite{allen}), using the Kubo formula:
\begin{eqnarray}
\sigma_{2D}^{xx}(\omega) &=& \frac{\sigma_{0}}{N}\sum_{\alpha,\beta}
{ {n_{\alpha}-n_{\beta}} \over {\epsilon_{\beta}-\epsilon_{\alpha}} } 
|\langle \alpha|J_x|\beta\rangle|^2
  \delta(\omega-(\epsilon_{\beta}-\epsilon_{\alpha}))
\nonumber
\end{eqnarray}
Where, the current operator $J_x$ is
\begin{eqnarray}
J_x &=& -i\sum_{i,\sigma}\left[t(c^{\dagger}_{i,\sigma}c_{i+\hat{x},\sigma}-\textrm{hc})+t'(c^{\dagger}_{i,\sigma}c_{i+\hat{x}+\hat{y},\sigma}-\textrm{hc})\right]
\nonumber
\end{eqnarray}

The d.c conductivity is the $\omega \rightarrow 0$ limit of 
the result above.
$\sigma_{0}$=$\frac{\pi e^2}{\hbar}$, the scale for
two dimensional conductivity, has the dimension of {\it
conductance}.
$n_{\alpha}=f(\epsilon_{\alpha})$ is the Fermi function,
and $\epsilon_{\alpha}$ and $|\alpha\rangle$ are 
respectively the single particle eigenvalues and eigenstates of
$H_{eff}$ in a given background \{${\bf m}_i$\}. 
The results we show in the text are 
averaged over equilibrium MC configurations.

The experimental results are quoted as {\it resistivity} of a three
dimensional material. If we assume that the planes are 
electronically decoupled, as we have done 
in the text, then the
three dimensional resistivity $\rho_{3D}$ can 
be estimated from the resistance of a cube of size $L^3$.
If the 2D resistivity is 
$\rho_{2D} = 1/\sigma_{2D}$, the
resistance of a $L^2$ sheet is just $\rho_{2D}$ itself.
A stacking of such sheets, with spacing $c_0$ in the
third direction, implies that the resistance of the
$L^3$ system would be $R_{3D} = \rho_{2D} c_0/L$. By
definition this also equals $\rho_{3D} L/L^2 = \rho_{3D}/L$.
Equating the two, $\rho_{3D} = \rho_{2D} c_0$. 

\begin{figure*}[t]
\centerline{
\includegraphics[width=16.0cm,height=11cm]{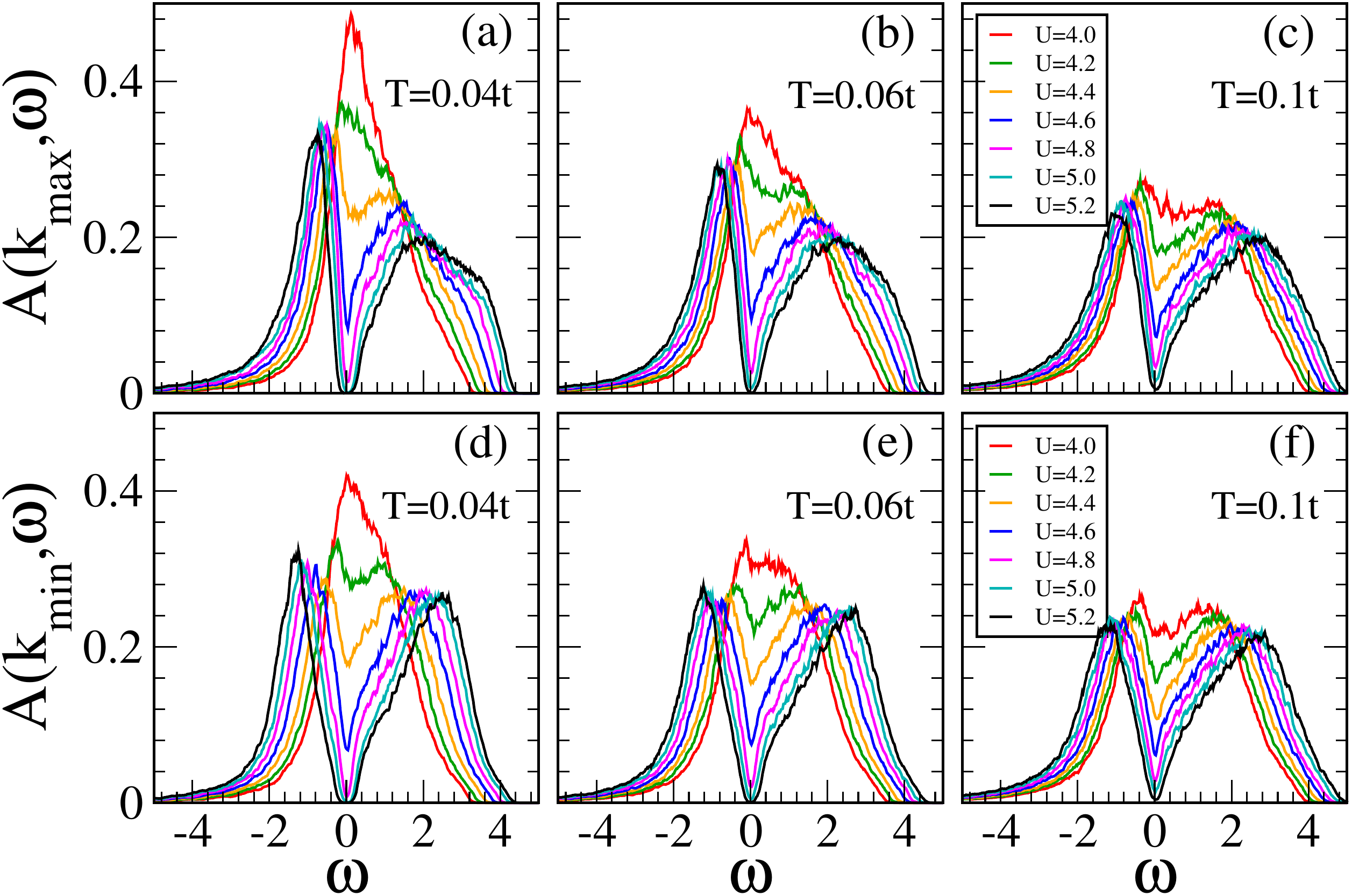}
}
\caption{\label{akf}The spectral function $A({\bf k}, \omega)$ 
at two ${\bf k}$ points on the FS that
correspond to the highest and  lowest value of $A({\bf k},0)$.
We highlight the anisotropy over a range of $U/t$ values,
as the system evolves from a moderately damped metal to
a pseudogap phase, and three temperatures.
}
\end{figure*}

\subsection{Spectral function}
We extract the thermal and spin averaged 
spectral function $A({\bf k},\omega)$ as 
follows.
First,  the retarded Greens function 
$$
G_{\sigma}({\bf k},t)=-i\theta(t)\langle
\{c_{{\bf k}\sigma}(t),c^{\dagger}_{{\bf k}\sigma}(0)\}\rangle
$$
 which can be simplified to
$$
G_{\sigma}({\bf k},t)=-i\theta(t)\sum_{\alpha}
|\langle {\bf k}\sigma|\alpha \rangle|^2e^{-i\epsilon_{\alpha}t}
$$
where $\{|\alpha\rangle\}$ are the 
single particle 
eigenstates and $\epsilon_{\alpha}$ 
are eigenvalues 
in a given $\{ {\bf m}_i \}$ background.
In frequency domain, this  becomes
$$
G_{\sigma}({\bf k},\omega)=
\sum_{\alpha}\frac{|\langle {\bf k}\sigma|\alpha \rangle|^2}{
\omega-\epsilon_{\alpha}+i0^+}
$$
From this:
$A_{\sigma}({\bf k},\omega)=-{1 \over \pi}$ 
Im$G_{\sigma}({\bf k},\omega)$ 
is simply 
$\sum_{\alpha}|\langle {\bf k}\sigma|\alpha \rangle|^2
\delta(\omega-\epsilon_{\alpha})$. We average this over
spin orientations, $\sigma$, and over thermal configurations.
The ${\bf k}$ dependent weight at $\omega=0$ is shown 
in Fig.5 in the text. The full spectral function at the
`cold spot' and `hot spot', where $A({\bf k},0$ is
maximum and minimum,  are shown, respectively, in
the top and bottom panels in the figure \ref{akf}.

We have averaged the spectrum over the four ${\bf k}$
neighbours of the nominal `cold' and `hot' points of
our $24 \times 24$ lattice. This averaging {\it reduces}
the anisotropy, so the true anisotropy would be greater
than what we show here. Also note that at $T=0.1t$, where
the $A({\bf k},0)$ in the text is shown, the spectral
function has no {\it peak} at $\omega=0$ either in the
cold or hot spot. The pseudogap feature is visible 
all over the FS even at $U/t=4$.

\bibliographystyle{unsrt}

\end{document}